\documentclass[pre,twocolumn,floatfix,aps,showpacs]{revtex4}
\usepackage{amssymb}
\usepackage{graphicx}
\usepackage{epsf}
\usepackage{amsmath}
\usepackage{bm} 
\usepackage{pspicture}
\usepackage{flafter}
\usepackage{float}
\begin{document}

\title{Fluctuation theorem for entropy production during effusion of an ideal gas with momentum transfer}
\author{Kevin Wood$^{1,2}$, C. Van den Broeck$^{3}$, R. Kawai$^{4}$,
and Katja Lindenberg$^{1}$}
\affiliation{
$^{(1)}$Department of Chemistry and Biochemistry and Institute for
Nonlinear Science, and $^{(2)}$ Department of Physics,
University of California San Diego, 9500 Gilman Drive,
La Jolla, CA 92093-0340, USA\\
$^{(3)}$Hasselt University, Diepenbeek, B-3590 Belgium\\
$^{(4)}$ Department of Physics, University of Alabama at Birmingham,
Birmingham, AL 35294 USA
}
\date{\today}

\begin{abstract}
We derive an exact expression for entropy production
during effusion of an ideal gas driven by momentum transfer in
addition to energy and particle flux. Following the treatment in
Phys. Rev. E {\bf 74}, 021117 (2006), we construct a master
equation formulation of the
process and explicitly verify the thermodynamic fluctuation theorem,
thereby directly exhibiting its extended applicability to particle
flows and hence to hydrodynamic systems.
\end{abstract}
\pacs{05.70.Ln, 05.40.-a, 05.20.-y}

\maketitle

\section{Introduction}

Since the pioneering work of Onsager~\cite{onsager} on the
relation between linear response and equilibrium fluctuations, his
insights have been further formalized in, for example, the theory of
linear irreversible processes \cite{prigogine} and the
fluctuation-dissipation theorem~\cite{callen}. Over the past decade some
new surprising results have been discovered that
suggest relations valid far away from equilibrium, notably  the
fluctuation~\cite{fluctuation} and work~\cite{work} theorems. The
fluctuation theorem, originally demonstrated for nonequilibrium
steady states in thermostated systems, has been proven in a  number
of  different settings. Basically, it states that during an experiment
of duration $t$, it is exponentially more likely to observe a positive
entropy production $\Delta S$ rather than an equally large negative one,
\begin{equation}
\label{flucthm}
\frac{P_t(\Delta S)}{{P_t}(-\Delta S)}=e^{\Delta S / k}.
\end{equation}
In the application to nonequilibrium steady states, the above
result is typically only valid in the asymptotic limit
$t \to \infty$ and expresses a symmetry property of large deviations.

We address another scenario in which the system is perturbed out of a
state which is initially at equilibrium. The so-called transient
fluctuation theorem is then valid for all times $t$.  We consider
the problem of a Knudsen flow between ideal gases that have overall
non-zero momentum. In this case, the stationary state is reached
instantaneously, so that there is no distinction between the transient
and steady state versions of the theorem. We show that the system
obeys a detailed fluctuation theorem which includes (\ref{flucthm})
as a special case.  Our calculation is an extension  of the one given
in~\cite{cleuren} to include momentum transfer. The interest of this
extension is manifold. First, the derivation of the fluctuation
theorem is somewhat more complicated since the momentum is a quantity
which is odd under velocity inversion. Second, momentum, together with
particle number and energy, are the conserved quantities whose
transport forms the basis of hydrodynamics. Our derivation therefore
puts the fluctuation theorem fully in this context (see also~\cite{bl}). 
Finally, as a bi-product, we calculate the Onsager matrix for the
Knudsen flow problem including momentum transport.

In Sec.~\ref{theorem} we formulate the fluctuation theorem for effusion
with momentum transfer for an ideal gas.  Section~\ref{master}
generalizes the derivation of the master equation and cumulant
generating function in~\cite{cleuren} to the case with momentum
transfer. Verification of the fluctuation theorem is detailed in
Sec.~\ref{verify}, and the lowest order cumulants are exhibited in
Sec.\ref{cumulants}.  We use these results to verify the Onsager
relations for this nonequilibrium system in Sec.\ref{onsager}.
We end with a brief conclusion in Sec.~\ref{conclusion}.  Some details
of the calculations are presented in appendices.

\begin{figure}
\includegraphics[width = 7cm]{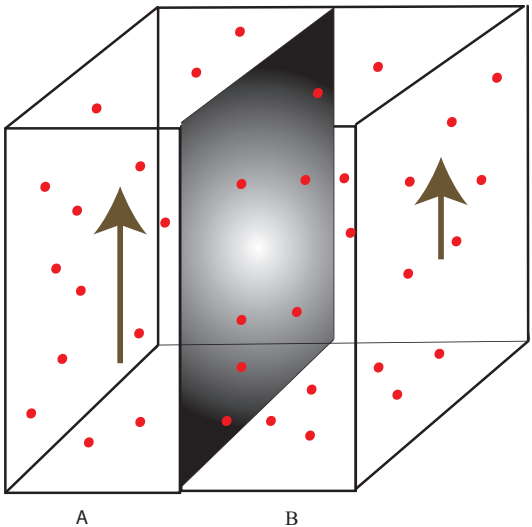}
\caption{The ideal gas in compartment $A$ is characterized by
temperaturer$T_A$, number density $n_A$, and flow velocity $V_A$. 
Similarly, the ideal gas in compartment $B$ is characterized by
temperature $T_B$, number density $n_B$, and flow velocity $V_B$. 
Velocities $V_A$ and $V_B$ are represented by the vertical
arrows within each compartment and are taken to be parallel to
the adiabatic wall separating the the two gases.}
\label{Fig1}
\end{figure}

\section{Fluctuation Theorem for Effusion  with Momentum Transfer}
\label{theorem}
We begin by considering two (infinitely) large neighboring reservoirs,
$A$ and $B$, each of which contains an ideal gas of uniform density $n_i$
in equilibrium at temperature $T_i$, $i \in {A,B}$.  In addition, the
particles of gas $i$ have an overall center of mass velocity $V_{i}$
in the ${\bf e_x}$ direction (Fig.~\ref{Fig1}). That is, the
velocity distributions of the gas particles take the Maxwellian
form
\begin{equation}
\phi_{i} = \left(\frac{m}{2 \pi k T_i} \right)^{3/2}
\exp\left({\frac{m ({\bf v} - V_{i} {\bf e_x})^2}{2 k T_i}}\right).
\label{maxwellian}
\end{equation}
The two reservoirs are separated by a common adiabatic wall parallel to
the ${\bf e_x}$ direction, with a hole of surface area $\sigma$ whose
linear dimensions are small compared with the mean free path of the
particles.  As a result, the local equilibrium in each reservoir is
not disturbed by the exchange of mass, heat, and momentum during a
finite time interval $t$ in which the hole is open. Upon a transfer
of total energy $\Delta U$,
particles $\Delta N$, and momentum $\Delta p_x$ during this time
interval, the overall change in entropy for the system is given by
\begin{equation}
\label{dS}
\begin{aligned}
\Delta S &= \Delta S_A + \Delta S_B \\
     &= -\frac{1}{T_A} \Delta U + (\frac{\mu_A}{T_A}-\frac{m V_A^2}
{2 T_A})\Delta N + \frac{V_A}{T_A} \Delta p_x  \\
 &~~~+ \frac{1}{T_B} \Delta U - (\frac{\mu_B}{T_B}-\frac{m V_B^2}{2 T_B})
\Delta N - \frac{V_B}{T_B} \Delta p_x \\
     &= \mathcal{A}_U \Delta U + \mathcal{A}_N \Delta N
+ \mathcal{A}_{p_x} \Delta p_x,
\end{aligned}
\end{equation}
where we have introduced the thermodynamic forces for energy,
particle, and momentum transfer~\cite{kitahara}:
\begin{equation}
\label{forces}
\begin{aligned}
\mathcal{A}_U &= \frac{1}{T_B} - \frac{1}{T_A}, \\
\mathcal{A}_N &= \frac{\mu_A}{T_A}-\frac{m V_A^2}{2 T_A} - (\frac{\mu_B}{T_B}-\frac{m V_B^2}{2 T_B})  \\
 &= k \log \bigg\lbrack \frac{n_A}{n_B} \biggl(\frac{T_B}{T_A}
\biggr)^{\frac{3}{2}} \bigg\rbrack +
(\frac{m V_B^2}{2 T_B}-\frac{m V_A^2}{2 T_A}), \\
\mathcal{A}_{p_x} &= \frac{V_A}{T_A} - \frac{V_B}{T_B}.
\end{aligned}
\end{equation}
In the equation for $\mathcal{A}_N$, we have used the expression for
the chemical potential $\mu$ of an ideal gas at rest.
Since the explicit expression for the thermodynamic forces in systems
with momentum is not readily available, we provide a brief derivation
in Appendix~\ref{appa}.

The variables $\Delta U$, $\Delta N$, and $\Delta p_x$ all
correspond to fluctuating quantities influenced by  single
particle crossings on each side of the adiabatic wall.  As a result,
the total entropy production $\Delta S$, which will be observed during
the time duration $t$, is likewise a fluctuating
quantity. However, time reversal symmetry of the microscopic dynamics
a relation for the probability distribution of this entropy production,
as expressed in Eq.~(\ref{flucthm}). 
Due to the absence of memory effects,
$\Delta S$ is in fact a stochastic process with independent
increments:  contributions to $\Delta
S$ from any two equal, non-overlapping time intervals are
independent identically distributed
random variables. It is therefore convenient to introduce the 
cumulant generating function, which takes the
form
\begin{equation}
\langle e^{- \lambda \Delta S} \rangle = e^{{-t g(\lambda)}}.
\end{equation}

The fluctuation theorem, Eq.~\ref{flucthm}, implies the following
symmetry property:
\begin{equation}
\label{symrel}
g(\lambda) = g(k^{-1} - \lambda).
\end{equation}
As the derivation in Appendix~\ref{DetailFT} points out, one can augment
the observation of the entropy production with additional variables,
while retaining the form of the fluctuation theorem. Hence the following
more detailed
fluctuation theorem, which is expressed in terms of the joint
probability density involving all three conserved quantities, particle
number, momentum and energy, is obtained:
\begin{equation}
\frac{P_t(\Delta U,\Delta N, \Delta p_x)}{{P}_t
(-\Delta U,-\Delta N,-\Delta p_x)}=e^{\Delta S / k}.
\label{detailflucthm}
\end{equation}
Since the increments of $\Delta U$, $\Delta
N$, and $\Delta p_x$ are also independent, we can write the
corresponding cumulant generating function as
\begin{equation}
\langle e^{-( \lambda_U \Delta U+\lambda_N \Delta N + \lambda_{p_x}
\Delta p_x)} \rangle = e^{-t g(\lambda_U,\lambda_N,\lambda_{p_x})}.
\label{mudef}
\end{equation}
The detailed fluctuation theorem then requires the following
symmetry relation, similar to Eq.~(\ref{symrel}):
\begin{equation}
\begin{aligned}
g(\lambda_U,&\lambda_N,\lambda_{p_x})  \\ &=g(\mathcal{A}_U k^{-1}
- \lambda_U,\mathcal{A}_N k^{-1} - \lambda_N,\mathcal{A}_{p_x} k^{-1}
- \lambda_{p_x}).
\end{aligned}
\label{musymdetail}
\end{equation}

Note finally that  Eq.~(\ref{detailflucthm}), apart from implying the
normal fluctuation theorem Eq.~(\ref{flucthm}), also implies fluctuation
theorems for particle,
energy, and momentum transfer individually when the complementary
thermodynamic forces are zero:

\begin{equation}
\begin{aligned}
\frac{\mathcal{P}_t(\Delta U)}{\mathcal{P}_t(-\Delta U)} &=
e^{\Delta S / k}; \phantom{+} \mathcal{A}_N=\mathcal{A}_{p_x} = 0,
\\
\frac{\mathcal{P}_t(\Delta N)}{\mathcal{P}_t(-\Delta N)} &=
e^{\Delta S / k}; \phantom{+} \mathcal{A}_U=\mathcal{A}_{p_x} = 0,
\\
\frac{\mathcal{P}_t(\Delta p_x)}{\mathcal{{P}}_t(-\Delta p_x)}
&=e^{\Delta S / k}; \phantom{+} \mathcal{A}_U=\mathcal{A}_{N} = 0.
\\
\end{aligned}
\end{equation}

\section{Master Equation and Cumulant Generating Function}
\label{master}
If we choose a sufficiently small time interval $dt$, the
contributions to the quantities $\Delta U$, $\Delta N$, and $\Delta
p_x$ arise from individual particles crossing the hole.  The kinetic
theory of gases allows us to calculate the probability per unit
time, $T_{i \to j}(E,p_x)$, to observe a particle with
kinetic energy $E = \frac{1}{2} m v^2$ and momentum $p_x = m v_x$
crossing the hole from reservoir $i$ to reservoir $j$.
Specifically, the transition rate in question is given by (see
Appendix~\ref{appa}):
\begin{equation}
\begin{aligned}
T_{i\to j}(&E,p_x) = \frac{\sigma n_i}
{m (k \pi T_i)^{3/2}} \left(E-\frac{\displaystyle p_x^2}
{\displaystyle 2 m}\right)^{1/2}\\
&\times\exp\left[- \frac{m}{2 k T_i} \left(
\frac{2(E-\frac{p_x^2}{2 m})}{m} + (\frac{p_x}{m} + V_{i})^2 \right)
\right],
\end{aligned}
\label{transrates}
\end{equation}
with $(i,j) = (A,B)$ or $(i,j) = (B,A)$. Hence,
the probability density $P_t(\Delta U, \Delta N,\Delta p_x)$ obeys
the master equation
\begin{equation}
\begin{aligned}
\partial_tP_t(&\Delta U, \Delta N,\Delta p_x)  \\
=&
\int_{-\infty}^{\infty} dp_x \int_{p_x^2/2 m}^{\infty} dE\;
T_{A \to B}\\
&~~\times P_t(\Delta U-E, \Delta N-1,\Delta p_x-p_x) \\
+& \int_{-\infty}^{\infty} dp_x \int_{p_x^2/2 m}^{\infty} dE\;
T_{B \to A} \\
&~~\times P_t(\Delta U+E, \Delta N+1,\Delta p_x+p_x) \\
-&P_t (\Delta U, \Delta N,\Delta p_x) \int_{-\infty}^{\infty}
dp_x \int_{p_x^2/2 m}^{\infty} dE\; \\
&~~\times \left( T_{A \to B} + T_{B \to A} \right).
\end{aligned}
\end{equation}
We have written $T_{i \to j}(E,p_x)$ without the arguments
for economy of notation.
We can take advantage of the convolution structure of the integral
operators by considering the equation in Fourier space; that is,
we multiply both sides of the equation by
$\exp[- (\lambda_U \Delta U + \lambda_N \Delta N +
\lambda_{p_x} \Delta p_x)]$ and integrate $\Delta U$, $\Delta p_x$
over all space and sum over all integers $\Delta N$.  We arrive at the
expression
\begin{equation}
\begin{aligned}
\partial_t \tilde{P}(&\lambda_U,\lambda_N,\lambda_{p_x}) = \tilde{P}
(\lambda_U,\lambda_N,\lambda_{p_x}) I_1 \\ &+
\tilde{P}(\lambda_U,\lambda_N,\lambda_{p_x}) I_2
- \tilde{P}(\lambda_U,\lambda_N,\lambda_{p_x}) I_3,
\end{aligned}
\end{equation}
where
\begin{equation}
\begin{aligned}
I_1 &= e^{- \lambda_N} \int_{-\infty}^{\infty} dp_x
\int_{p_x^2/2 m}^{\infty} dE\; T_{A \to B}
e^{-(\lambda_U E+\lambda_{p_x} p_x)}, \\
I_2 &= e^{\lambda_N} \int_{-\infty}^{\infty} dp_x
\int_{p_x^2/2 m}^{\infty} dE\; T_{B \to A}
e^{(\lambda_U E+\lambda_{p_x} p_x)}, \\
I_3 &= \int_{-\infty}^{\infty} dp_x \int_{p_x^2/2 m}^{\infty}
dE \; \left(T_{A \to B} + T_{B \to A}\right).
\label{integrals}
\end{aligned}
\end{equation}

From this expression we can write 
$g(\lambda_U,\lambda_N,\lambda_{p_x})$, defined in Eq.~(\ref{mudef}),
as
\begin{equation}
g(\lambda_U,\lambda_N,\lambda_{p_x}) = I_3 - (I_1 + I_2).
\end{equation}
The integrals (\ref{integrals}) can easily be performed by switching
to the variable $z = E-p_x^2/2 m$ and integrating $z$ from
zero to infinity and $p_x$ over all space, as before.  We thereby
arrive at our final expression for
$g(\lambda_U,\lambda_N,\lambda_{p_x})$:
\begin{equation}
\begin{aligned}
g(\lambda_U,&\lambda_N,\lambda_{p_x})= \sigma \left(\frac{k}{2\pi m}
\right)^{1/2} \\
&\times \left[ n_A T_A^{1/2}
\left(1 - \frac {G_A}{(1+kT_A\lambda_U)^2}\right)\right.\\
&~~~+ \left.n_B T_B^{1/2}
\left(1 - \frac {G_B}{(1-kT_B\lambda_U)^2}\right)\right],
\end{aligned}
\end{equation}
where
\begin{equation}
\begin{aligned}
G_A&\equiv \exp\left( - \lambda_N - \frac{m V_A^2 \lambda_U
- k m T_A \lambda_{p_x}^2 + 2 m  V_A \lambda_{p_x}}
{2 (1 + k T_A \lambda_U)}\right),\\
G_B&\equiv \exp\left(\lambda_N + \frac{m V_B^2 \lambda_U
+ k m T_B \lambda_{p_x}^2 + 2 m  V_B \lambda_{p_x}}
{2 (1 - k T_B \lambda_U)}\right).
\end{aligned}
\end{equation}
Notice that $g(\lambda_U,\lambda_N,\lambda_{p_x})$ can be written as
a sum of two contributions,
\begin{equation}
g(\lambda_U,\lambda_N,\lambda_{p_x})=g_A(\lambda_U,\lambda_N,\lambda_{p_x})
+g_B(\lambda_U,\lambda_N,\lambda_{p_x}),
\end{equation}
with
\begin{equation}
\begin{aligned}
g_A(\lambda_U,\lambda_N,\lambda_{p_x})&=
\sigma \left(\frac{k}{2 \pi m}\right)^{1/2}\\
&\times 
n_AT_A^{1/2} 
\left(1 - \frac{G_A}{(1+kT_A\lambda_U)^2}\right),\\
g_B(\lambda_U,\lambda_N,\lambda_{p_x})&=
\sigma \left(\frac{k}{2 \pi m}\right)^{1/2}\\
&\times 
n_BT_B^{1/2} 
\left(1 - \frac{G_B}{(1-kT_B\lambda_U)^2}\right).
\end{aligned}
\end{equation}
This additivity
property arises from the statistical independence of the fluxes from
$A \to B$ and $B \to A$.

\section{Fluctuation Symmetry}
\label{verify}
We now proceed to explicitly verify the symmetry relation~(\ref{musymdetail})
and hence the fluctuation theorem.
Conceptually, one may understand this symmetry relation as follows.
Under the symmetry operation $\mathcal{T}$ -- that is, under the
transformation given by the r.h.s. of Eq.~(\ref{musymdetail}) --
the term containing the exponential in $g_A$ (which we call
$g_{A,1}$) becomes the corresponding term from $g_B$ (which we
call $g_{B,1}$) and similarly, the original $g_{B,1}$ term
becomes $g_{A,1}$, thereby preserving the overall structure of
$g$. Mathematically, we can express this as:
\begin{equation}
\begin{aligned}
\mathcal{T}[g_{A,1}] &= g_{B,1}, \\
\mathcal{T}[g_{B,1}] &= g_{A,1},
\end{aligned}
\end{equation}
where
\begin{equation}
\label{muaexponent}
g_{A,1} = \sigma \left({k}{2 \pi m}\right)^{1/2} n_A T_A^{1/2}(1+k
T_A \lambda_U)^{-2} G_A,
\end{equation}
and
\begin{equation}
\label{mubexponent}
g_{B,1} = \sigma\left({k}{2 \pi m}\right)^{1/2} n_B T_B^{1/2}(1-k
T_B \lambda_U)^{-2} G_B.
\end{equation}
Explicitly, we have
\begin{equation}
\begin{aligned}
\mathcal{T}[&g_{A,1}]=\sigma \left(\frac{k}{2 \pi m}\right)^{1/2}
n_A T_A^{1/2} \left(1+k T_A \Lambda_U\right)^{-2} \\
&\times
\exp\left\{\log \left[\frac{n_A}{n_B} \left(\frac{T_B}{T_A}
\right)^{3/2}\right]  +\Lambda_N \right\} \\ &\times
\exp \left[
\frac{-m V_A^2 \Lambda_U + k m T_A \Lambda_{p_x}^2}
{2 (1 + k T_A \Lambda_U)}\right] \\ &\times
\exp\left( \frac{- 2 m  V_A \lambda_{p_x}}{2 (1 + k T_A
\Lambda_U)}\right) \\ &=
\sigma \left(\frac{k}{2 \pi m}\right)^{1/2} n_A T_A^{1/2} 
\Lambda^{-2} \left(\frac{n_A}{n_B}
\frac{T_B}{T_A}\right) \\ &\times
\exp\left[-\lambda_N 
+ \frac{2\Lambda
\left(\frac{\displaystyle V_B^2
m}{2kT_B} -\frac{\displaystyle V_A^2 m}{\displaystyle 2 k T_A}
\right) -m V_A^2 \Lambda_U}{2\Lambda} \right]
\\ &\times
\exp\left(\frac{kmT_A\Lambda_{p_x}^2 - 2 m  V_A \lambda_{p_x}}{2\Lambda}
\right),
\end{aligned}
\end{equation}
where
\begin{equation}
\begin{aligned}
\Lambda_U&\equiv \frac{1}{kT_B} -\frac{1}{kT_A} -\lambda_U,\\
\Lambda_{p_x} &\equiv \frac{V_A}{kT_A} - \frac{V_B}{kT_B}
-\lambda_{p_x},\\
\Lambda_N &\equiv \frac{V_B^2m}{2kT_B}-\frac{V_A^2m}{2kT_A} -\lambda_N,\\
\Lambda&\equiv \frac{T_A}{T_B}\left( 1-kT_B\lambda_U\right).
\end{aligned}
\end{equation}
Following simplification, this reduces to $g_{B,1}$.  A similar
result holds for $\mathcal{T}[g_{B,1}]$, and therefore the
fluctuation theorem symmetry is satisfied.

\section{Cumulants}
\label{cumulants}
The joint cumulant $\kappa_{ijk}$ of power i in energy flux, j in
particle flux, and k in momentum flux appears as a coefficient in
the Taylor expansion of the cumulant generating function, namely,
\begin{equation}
g_A(\lambda_U,\lambda_N,\lambda_{p_x})=
- \frac{1}{t} \sum_{i,j,k=0}^\infty \frac{(-1)^{i+j+k}
\lambda_U^i \lambda_N^j \lambda_{p_x}^k}{i!j!k!} \kappa_{ijk} .
\end{equation}
While our expression for $g_A(\lambda_U,\lambda_N,\lambda_{p_x})$
allows us to calculate joint cumulants of any order, we here mention
only the first order results, which are relevant for verifying the
Onsager relations in the subsequent section:
\begin{equation}
\begin{aligned}
\kappa_{100} &= \langle \Delta U \rangle   \\
&=t \sigma \left(\frac{k}{2 \pi m}\right)^{1/2}
\left( n_A T_A^{1/2}(2 k T_A + \frac{m V_A^2}{2}) \right.\\
&~~~ \left.  - n_B T_B^{1/2}(2 k T_B + \frac{m V_B^2}{2}) \right),  \\
\kappa_{010} &= \langle \Delta N \rangle \\
&=  t \sigma \left(\frac{k}{2
\pi m}\right)^{1/2} \left( n_A T_A^{1/2} - n_B T_B^{1/2} \right), \\
\kappa_{001} &= \langle \Delta p_x \rangle \\
& =   t \sigma 
\left(\frac{k m}{2 \pi}\right)^{1/2} \left(n_A V_AT_A^{1/2}-
n_B V_B T_B^{1/2} \right).
\label{means}
\end{aligned}
\end{equation}
Note that the cumulant associated with energy $\langle \Delta U
\rangle$ contains terms corresponding to both particle transport and
momentum transport.

\section{Onsager Relations}
\label{onsager}
Averaging Eq.~(\ref{dS}) and taking the time derivative leads
us to an equation for the average entropy production,
\begin{equation}
\frac{d}{dt} \langle \Delta S \rangle = \mathcal{J}_U
\mathcal{A}_U + \mathcal{J}_N \mathcal{A}_N+
\mathcal{J}_{p_x} \mathcal{A}_{p_x},
\end{equation}
with the macroscopic fluxes $\mathcal{J}_X$ defined as
\begin{equation}
\label{fluxes}
\begin{aligned}
\mathcal{J}_U &= \frac{d}{dt} \langle \Delta U \rangle  \\
&= \sigma \left(\frac{k}{2 \pi m}\right)^{1/2} \left( n_A T_A^{1/2}(2 k T_A
+ \frac{m V_A^2}{2})\right.\\
&~~~-\left.  n_B T_B^{1/2}(2 k T_B + \frac{m V_B^2}{2}) \right),  \\
\mathcal{J}_N &= \frac{d}{dt} \langle \Delta N \rangle =  \sigma
\left(\frac{k}{2 \pi m}\right)^{1/2} \left( n_A T_A^{1/2}
- n_B T_B^{1/2} \right), \\
\mathcal{J}_{p_x} &= \frac{d}{dt} \langle \Delta p_x \rangle \\
&=  
\sigma \left(\frac{k m}{2 \pi}\right)^{1/2} \left( n_A V_AT_A^{1/2}-
n_B V_B T_B^{1/2} \right).
\end{aligned}
\end{equation}
While these fluxes are in general complicated nonlinear functions of
the affinities
($\mathcal{A}_U$,$\mathcal{A}_N$,$\mathcal{A}_{p_x}$), near
equilibrium we can write:
\begin{equation}
\begin{aligned}
T_A & = T - \frac{\Delta T}{2}, \;\;\;\;
&T_B  = T + \frac{\Delta T}{2}, \\
n_A & = n - \frac{\Delta n}{2}, \;\;\;\;
&n_B  = n + \frac{\Delta n}{2}, \\
V_A &= V - \frac{\Delta V}{2}, \;\;\;\;
&V_B = V + \frac{\Delta V}{2},
\end{aligned}
\end{equation}
and expand the forces and fluxes to first order in the small deviations
$\Delta T$, $\Delta n$, and $\Delta V$.  To linear order, the
thermodynamic forces become
\begin{equation}
\begin{aligned}
\mathcal{A}_U &= -\frac{\Delta T}{T^2}; \\
\mathcal{A}_N &= \frac{m V}{T} \Delta V + (\frac{3 k}{2 T}
- \frac{m V^2}{2 T^2})\Delta T  -  \frac{k}{n}\Delta n \\
\mathcal{A}_{p_x} &= - \frac{\Delta V}{T} +  \frac{V}{T^2}\Delta T.
\end{aligned}
\end{equation}
Taylor expansions of the fluxes $\mathcal{J}_U$,$\mathcal{J}_N$,and
$\mathcal{J}_{p_x}$ [Eq.~(\ref{fluxes})] allow us to write
\begin{equation}
\label{onsagrel}
\bar{J} = \bm{O} \bar{A},
\end{equation}
where $\bar{J} = (\mathcal{J}_U,\mathcal{J}_N,\mathcal{J}_{p_x})^T$ and
$\bar{A} = (\mathcal{A}_U,\mathcal{A}_N,\mathcal{A}_{p_x})^T$.  The
Onsager matrix $\bm{O}$ is given by:
\begin{widetext}
\begin{equation}
\bm{O} = \sigma \left(\frac{k}{2\pi m}\right)^{1/2} n T^{3/2}
\begin{pmatrix}
6 k T + 3 m V^2 + \frac{\left(mV^2\right)^2}{4kT}& 2 + \frac{m V^2}{2kT} &
~~\frac{1}{2}mV(6+\frac{mV^2}{kT})~~ \\ \\
2 + \frac{m V^2}{2kT} & \frac{1}{k T} & ~~\frac{mV}{kT}~~ \\ \\
\frac{1}{2}mV(6+\frac{mV^2}{kT}) & ~~\frac{mV}{kT}~~ & m +
\frac{(mV)^2}{kT}
\end{pmatrix},
\end{equation}
\end{widetext}
which clearly has the required symmetry $O_{ij}=O_{ji}$.  
The Onsager relations [Eq.~(\ref{onsagrel})] fully detail the complex
coupling between energy, particle, and momentum transport in the linear
regime.  

Note that in the case of moving gases, $V \neq 0$, 
the presence of a temperature gradient alone
($\Delta n= \Delta V = 0$) is sufficient to produce a nonzero net
flux of momentum.  Note also that
when there is only a momentum gradient, $\Delta T = \Delta n = 0$,
the heat, particle, and momentum fluxes reduce to:
\begin{equation}
\begin{aligned}
\mathcal{J}_U &= - \frac{\sigma k^{\frac{1}{2}} n T^{\frac{1}{2}}}{(2 \pi
m)^{\frac{1}{2}}} (m V) \Delta V, \\ 
\mathcal{J}_N &= 0, \\
\mathcal{J}_p &= - \frac{\sigma k^{\frac{1}{2}} n T^{\frac{1}{2}}}{(2 \pi m)^{\frac{1}{2}}} m \Delta V.  
\end{aligned}
\end{equation}
Therefore, when we choose the velocities to be equal but opposite so that $V=0$, the only nonzero flux is due to momentum transport.  In other words, momentum exchange takes place without a net exchange of particles or energy.

\section{Conclusion}
\label{conclusion}
The work and fluctuation theorems are quite remarkable. They are
basically one further step in Onsager's program to  take into account
the time-reversal symmetry of the microscopic dynamics. This results in
a stringent constraint on the probability density of the entropy
production.  The implications of this result are still being explored.
In this paper we have shown by an explicit microscopically exact
calculation that the fluctuation theorem applies for the effusion
between ideal gases with non-zero overall momentum. This sets the
stage for the application of the formalism in fluctuating hydrodynamics.

\section*{Acknowledgments}
This work was partially supported by the National Science Foundation
under Grant No. PHY-0354937.

\appendix
\section{}
\label{appa}

We can derive the thermodynamic forces for an ideal gas of $N$ particles
in volume $\cal{V}$ with non-zero momentum by considering the
entropy $S(U,N,\cal{V})$ of a gas at rest as a function of $U$, the
total energy, $N$, and $\cal{V}$.
Because adding an overall velocity to the gas does not change its volume
in phase space and hence its entropy, we can write  the entropy
$S(U,N,{\cal{V}},p)$
of a flowing gas which depends on momentum $p$ in terms of the
entropy  $S_0(U,N,\cal{V})$ of a gas at rest:
\begin{equation}
\label{Ssym}
S(U,N,{\cal{V}},p)=S_0((U-\frac{p^2}{2 N m}),N,{\cal{V}})
= S_0(\epsilon,N,{\cal{V}}),
\end{equation}
where $\epsilon\equiv U-p^2/2Nm$ represents the internal energy of the
gas.  The Sackur-Tetrode formula~\cite{saktet} provides the
explicit expression for
$S(\epsilon,N,\cal{V})$  which, with Eq.~(\ref{Ssym}) leads to
\begin{equation}
\label{saktet}
S=k N \log \left[
\frac{{\cal{V}}}{N}\left(\frac{U-\frac{p^2}{2 m N}}{N}\right)^{3/2}
\right] + \frac{3}{2} k N \left[\frac{5}{3}+\log \left(
\frac{4 \pi m}{3 h^2} \right)\right].
\end{equation}
Here $h$ is Planck's constant and $m$ is the mass of a single gas particle.
We can write the total
entropy change of the effusion process considered here as
\begin{equation}
\begin{aligned}
dS &=\frac{\partial{S_A}}{\partial{U_A}} dU_A +
\frac{\partial{S_A}}{\partial{N_A}} dN_A +
\frac{\partial{S_A}}{\partial{p_A}}
dp_A \\ & + \frac{\partial{S_B}}{\partial{U_B}} dU_B +
\frac{\partial{S_B}}{\partial{N_B}} dN_B
+\frac{\partial{S_B}}{\partial{p_B}} dp_B \\ &= dU
(\frac{\partial{S_B}}{\partial{U_B}}-\frac{\partial{S_A}}{\partial{U_A}})+dN
(\frac{\partial{S_B}}{\partial{N_B}}-\frac{\partial{S_A}}{\partial{N_A}})
\\   & + dp (\frac{\partial{S_B}}{\partial{p_B}}-\frac{\partial{S_A}}
{\partial{p_A}}),
\end{aligned}
\end{equation}
where $S_i$ corresponds to Eq.~(\ref{saktet}) with $U \to U_i$,
$N \to N_i$, and $p \to p_i$, $i \in {A,B}$, and we have used
momentum, energy, and particle conservation to write
$dU=-dU_A=dU_B$, $dN=-dN_A=dN_B$, and $dp=-dp_A=dp_B$. Performing
the above calculations and considering that the total energy $U$ of
an ideal gas with overall momentum $p$ at temperature $T$ is given
by
\begin{equation}
U = \frac{3N}{2 k T} + \frac{p^2}{2 m },
\end{equation}
we arrive after simplification at the expressions given in
Eq.~(\ref{forces}).

\section{}
\label{DetailFT} 
We give a derivation of the fluctuation theorem,
Eq.~(\ref{detailflucthm}), by adapting to the present case
the procedure introduced in~\cite{cleurenprl}. We consider the
Hamiltonian evolution of a system, consisting of two disjoint
subsystems $A$ and $B$ initially at equilibrium characterized
by micro-canonical distributions  with total particle number, momentum,
energy and volume equal to $N_i$, $p_i$, $U_i$ and ${\cal{V}}_i$,
$i=A,B$, respectively. At the initial time, the constraint separating
both systems is broken. It is assumed that this can be achieved
without any external work, momentum exchange or other perturbation
of the subsystems. This is clearly the case for the opening of a
hole in the adiabatic wall separating  ideal gases, as considered here.
After  a  time interval of duration $t$, the constraint is again
introduced at no cost of energy or momentum. One records the new
values of the parameters $N'_i$, $p'_i$ and $U'_i$.
The  amounts $(\Delta N,\Delta p,\Delta U)$ that are transported from
system $A$ to system $B$ will depend on the specific run, i.e.,  on
the starting configuration at $t=0$.
Let the volume in phase space corresponding to the initial states that
lead to the transport of these amounts be denoted by
$\Omega_{(N_i, p_i ,U_i)}(\Delta N,\Delta p,\Delta U)$. 
The probability to observe
such a realization is then given by
\begin{equation}
\label{P1}
P_{(N_i, p_i ,U_i)}(\Delta N, \Delta p,\Delta U)
=\frac{\Omega_{(N_i, p_i ,U_i)}(\Delta N,\Delta p,\Delta U)}
{\Omega_{(N_i, p_i ,U_i)}},
\end{equation}
where $\Omega_{(N_i, p_i ,U_i)}$ is the total phase space volume.
We now apply this very same result for parameter values 
$N'_i$, $-p'_i$ and $U'_i$, $i=A,B$, and consider the probability of
transporting the amounts $(-\Delta N,\Delta p,-\Delta U)$. Apart from
velocity inversion, the final values in this procedure are then the
initial ones of the first scenario, i.e., $(N_i, -p_i ,U_i)$. The
corresponding probability reads
\begin{equation}
\label{P2}
\begin{aligned}
P_{(N'_i, -p'_i ,U'_i)}&(-\Delta N, \Delta p,-\Delta U)\\
&=\frac{\Omega_{(N'_i, -p'_i ,U'_i)}(-\Delta N,\Delta p,-\Delta U)}
{\Omega_{(N'_i, -p'_i ,U'_i)}}.
\end{aligned}
\end{equation}
By micro-reversibility, there is a one-to-one correspondence between
each trajectory in the first situation with the time reversed trajectory
in the second situation. Furthermore, since Hamiltonian dynamics
preserves phase volume, the numerators in the r.h.s. of
Eqs.~(\ref{P1}) and (\ref{P2}) are identical. We conclude that
\begin{eqnarray}\label{P3}
\frac{P_{(N_i, p_i ,U_i)}(\Delta N, \Delta p,\Delta U)}
{P_{(N'_i, -p'_i ,U'_i)}(-\Delta N, \Delta p,-\Delta U)}
&=&\frac{\Omega_{(N'_i, -p'_i ,U'_i)}}{\Omega_{(N_i, p_i ,U_i)}}
\nonumber \\
&=& \exp (\frac{\Delta S}{k_B}),
\end{eqnarray}
where we used the fact that the entropy of a state is Boltzmann's
constant time the logarithm of the phase space volume $\Omega$ of
that state. $\Delta S$ is thus the entropy difference between states with
and without the primes.
We now note that inverting the momenta of the gases does not change
the statistics of particle and energy transport, but will obviously
change the sign of the momentum transfer,
$P_{(N'_i, -p'_i ,U'_i)}(-\Delta N, \Delta p,-\Delta U)
=P_{(N'_i, p'_i ,U'_i)}(-\Delta N, -\Delta p,-\Delta U)$.
Hence we can rewrite Eq.~(\ref{P3}) as follows:
\begin{eqnarray}\label{P4}
\frac{P_{(N_i, p_i ,U_i)}(\Delta N, \Delta p,\Delta U)}{P_{(N'_i, p'_i ,U'_i)}(-\Delta N,- \Delta p,-\Delta U)}= \exp (\frac{\Delta S}{k})
\end{eqnarray}

Finally, we  consider the thermodynamic limit of infinitely large
systems  with finite particle density $n_i=N_i/{\cal{V}}_i$, momentum
density $m V_i=p_i/N_i$, and energy density $u_i=U_i/{\cal{V}}_i$.
We furthermore assume that the effects of the removal of the constraint do
not scale with the volume, so that during the finite time $t$ it
results in non-extensive changes in the parameter values.  Hence
we can drop the sub-indices of $P$ on the l.h.s. of Eq.~(\ref{P4}.
Adding the sub-index $t$ to emphasize the duration of the exchange,
one can thus rewrite Eq.~(\ref{P4}) as Eq.~(\ref{detailflucthm}) of the
main text.

\section{}
\label{appc}
Here we briefly derive the formula for the transition rate $T_{A \to
B}(E,p_x)$ using the kinetic theory of gases.  We consider the
$\hat{z}$ direction to point from reservoir $A$ to reservoir $B$. We
require $T_{A \to B}(E,p_x) dE dp_x dt$, the probability to
observe a particle with kinetic energy in the range $(E,E+dE)$ and
momentum in the range $(p_x,p_x + dp_x)$ to cross the hole from $A$
to $B$ in a time interval $dt$.  The $z$ component of the position
of a particle with velocity $\bm{v}$ must
be located within a cylinder of base area $\sigma$ (the area of the
hole) and height $v_zdt$ measured from the wall.  
Furthermore, it must be traveling in the
$+\bm{e_z}$ direction (towards the hole).  The appropriate expression
is
\begin{equation}
\begin{aligned}
 T&_{A \to B}(E,p_x) = \\ &\int_{-\infty}^{\infty} dv_x
\int_{-\infty}^{\infty} dv_y \int_{0}^{\infty} dv_z \sigma v_z n_A 
\phi_A(\bm{v},V_A) \\ & \times \delta(\frac{m v^2}{2} - E) \delta(m v_x - p_x),
\end{aligned}
\end{equation}
where $\phi_A(\bm{v},V_A)$ is the Maxwellian given by
Eq.~(\ref{maxwellian}), and we have explicitly noted its dependence on
$\bm{v}$ and $V_A$.  A similar equation holds for $T_{A \to B}$. 
The $v_x$ integral is trivial because of the second delta
function, and the remaining integrals can be easily performed by
changing to polar coordinates $(R,\theta)$, given by
\begin{equation}
R^2=v_y^2+v_z^2; \phantom{++}
\tan \theta = \frac{v_y}{v_z}.
\end{equation}
This yields the expressions given in Eq.~(\ref{transrates}).

\end{document}